\documentstyle[preprint,aps]{revtex}
\tightenlines
\begin{document}
\draft
\preprint{\today}
\title {Triaxial projected shell model approach \\ }

\author{J.A. Sheikh$^{(1,2,3)}$ and K. Hara$^{(1)}$}

\address{$^{(1)}$Physik-Department, Technische Universit\"at M\"unchen,
D-85747 Garching, Germany \\
$^{(2)}$Institut de Recherches Subatomiques (IReS),
Universite Louis Pasteur,
23 rue du Loess, F-67037 Strasbourg Cedex 2, France\\
$^{(3)}$Tata Institute of Fundamental 
Research, Colaba, Bombay - 400 005, India
}

\maketitle

\begin{abstract}
The projected shell model analysis is carried out using the triaxial
Nilsson+BCS basis. It is demonstrated that, for an accurate description
of the moments of inertia in the transitional region, it is necessary
to take the triaxiality into account and perform the three-dimensional
angular-momentum projection from the triaxial Nilsson+BCS intrinsic
wavefunction. 
\end{abstract}


The major advancement in the studies of deformed nuclei has been the
introduction of the Nilsson potential\cite{nil55}. It was shown that the
rotational properties of deformed nuclei can be described by considering
nucleons to move in a deformed potential. The description of the
deformed nuclei in medium and heavy mass regions is impossible using the
standard (spherical) shell model approach, despite the recent
progress in the computing power. The Nilsson model has provided a
useful nomenclature for the observed rotational bands. It is known that
each rotational band is built on an intrinsic Nilsson state. The Nilsson
or deformed state is defined in the intrinsic frame of reference in
which the rotational symmetry has been broken and in order to calculate
the observable properties, it is required to restore the broken symmetry.

The rotational symmetry can be restored by using the standard 
angular momentum projection operator \cite{rs80}. This method has been used
to project out the good angular momentum states from the Nilsson+BCS
intrinsic state \cite{hi79,hi80,hi82}, see also the review article
\cite{hs95} and references cited therein. In this approach, the angular
momentum projection is carried out from a chosen set of Nilsson+BCS
states near the Fermi-surface. The projected states are then used to
diagonalize a shell model Hamiltonian. This approach referred to as the
projected shell model (PSM) follows the basic philosophy of the standard
shell model approach. The only difference is that, in the PSM, the
deformed basis is employed rather than the spherical basis. This makes
the truncation of the many-body basis very efficient, so that the shell
model calculations even for heavier systems can be easily performed.

The PSM approach has been used to describe a broad range of nuclear
phenomena such as backbending \cite{hs90}, superdeformed
\cite{sun1,sun2} and identical bands \cite{sun3} with considerable
success. The assumption in the PSM approach has been the
axial symmetry for the deformed system to keep the computation simple.
In fact, this is a reasonable approximation for well deformed nuclei.
However, for transitional nuclei, this assumption is questionable. The
inadequacy of the axially symmetric basis has been clearly demonstrated
by moments of inertia (the backbending plots) of the transitional nuclei
in the rare-earth region. It has been shown \cite{hs95} that, in the low
spin region, observed moments of inertia for lighter rare-earth nuclei
(for instance $^{156}$Er, $^{158}$Er, $^{158}$Yb and $^{162}$Hf) and
for the heavier rare-earth nuclei (for instance $^{172}$W, $^{174}$W
and $^{176}$W) increase quite steeply with increasing rotational
frequency as compared to the moment of inertia calculated by
the axially symmetric PSM approach (see Figs. 14--17 in \cite{hs95}).
This can be understood by noting that the horizontal (vertical) line in
the backbending plot represents the rotational (vibrational) limit since
the energy $E(I)$ as a function of spin $I$ is proportional to $I^2$
($I$). In this sense, the experimental data slants towards the
vibrational side in comparison with the existing PSM results. On the
other hand, the spectrum of a triaxial rotor \cite{df58} is known to
vary from rotational spectrum to a vibrational one as the triaxiality
parameter $\gamma$ increases from 0 to $30^o$ and, using this model, it
has been demonstrated (see Fig. 18 in \cite{hs95}) that the backbending
plot indeed slants towards the vibrational limit when $\gamma$
increases. It is therefore expected that, by using the triaxial basis in
the PSM, the moments of inertia and other properties of the transitional
nuclei can be described more appropriately. As pointed out in \cite{hs95},
the major problem here lies rather in the ground state band. This part
of the spectrum is quite insensitive to the configuration mixing since
the energy and spin values are still low ($I \le 10$), so that an
improvement of the ground state by allowing some triaxiality is in
order. The purpose of the present work is to develop a triaxial
projected shell model (referred to as TPSM hereafter) approach for the
description of transitional nuclei. This requires a three-dimensional
angular momentum projection and has not been attempted so far except for
a short investigation in early eighties \cite{hhr84}. We have carried
out the three-dimensional projection and would like to report our
preliminary results.

The shell model Hamiltonian employed is identical to the one used in the
axially symmetric PSM approach \cite{hs95}. It consists of $Q \cdot Q$ +
monopole pairing + quadrupole pairing forces
\begin{equation}
\hat H = \hat H_0
-\frac{\chi}{2} \sum_{\mu} \hat Q^\dagger_\mu \hat Q^{}_\mu
-G_M \hat P^\dagger \hat P
-G_Q \sum_{\mu} \hat P^\dagger_\mu \hat P^{}_\mu.
\end{equation}
Here, $\hat H_0$ is the the spherical harmonic-oscillator
single-particle Hamiltonian with a proper $l \cdot s$-force while the
operators $\hat Q$ and $\hat P$ are defined as
\begin{equation}
\hat Q_\mu = \sum_{\alpha \beta} Q_{\mu \alpha \beta} c^\dagger_\alpha
c^{}_\beta,~~~
\hat P^\dagger = \frac{1}{2}\sum_{\alpha} c^\dagger_\alpha
c^\dagger_{\bar\alpha},~~~
\hat P^\dagger_\mu = \frac{1}{2}\sum_{\alpha \beta} Q_{\mu \alpha \beta} 
c^\dagger_\alpha c^\dagger_{\bar\beta},
\end{equation}
where the quadrupole matrix-elements are given by
\begin{equation}
Q_{\mu\alpha\alpha'} = \delta_{NN'} (Njm|Q_\mu|N'j'm').
\end{equation}
In Eq. (2), $\alpha = \{Njm\}$ while $\bar \alpha$ represents the
time-reversed state of $\alpha$. The Hartree-Fock-Bogoliubov (HFB)
approximation of the shell model Hamiltonian Eq. (1) leads to the
quadrupole mean-field which is similar to the Nilsson potential.
Therefore, instead of performing the HFB variational analysis of the
Hamiltonian in Eq. (1), the Nilsson  potential can be directly used to
obtain the deformed basis. In the present work, we use the triaxial
Nilsson potential specified by the deformation parameters $\epsilon$
and $\epsilon'$
\begin{equation}
\hat H_N= \hat H_0 - \frac{2}{3} \hbar\omega ( \epsilon \hat Q_0
+\epsilon' \frac{\hat Q_{+2}+\hat Q_{-2}}{\sqrt 2} ),
\end{equation}
to generate the deformed single-particle wavefunctions.
It can be easily seen that the rotation operator $e^{-\imath \frac{\pi}{2}
\hat J_z}$ transforms the Nilsson Hamiltonian $\hat H_N$ into the opposite
triaxiality $(\epsilon' \rightarrow -\epsilon')$ leaving the eigenvalues 
unchanged. Later, it will be shown that the projected energy is
independent of the sign of $\epsilon'$ and it is sufficient to consider
only the non-negative $\epsilon'$. The volume conservation also restricts
the range of $\epsilon$ and $\epsilon'$ values to
\begin{equation}
-3<\epsilon<{3 \over 2},~~~~|\epsilon'|<{\sqrt 3}(1+ {\epsilon \over 3}).
\end{equation}
The triaxial Nilsson potential has been solved for the rare-earth region
with three major shells $N=4, 5, 6~(3, 4, 5)$ for neutrons (protons).

In the next step, the monopole pairing Hamiltonian is treated based on
the triaxial Nilsson basis. We use the standard strengths for the
pairing interaction of the form
\begin{equation}
G_M = ( G_1 \mp G_2 { N-Z \over A}) {1 \over A} ,
\end{equation}
where $-$ $(+)$ is for neutrons (protons) while $G_1$ and $G_2$ are
chosen respectively as 21.24 and 13.86 MeV in the rare-earth region. The
pairing correlations are treated by using the usual BCS approximation
to establish the Nilsson+BCS basis. The three-dimensional angular
momentum projection is then carried out on the quasiparticle states
obtained in this way.

The three-dimensional angular momentum projection operator is given by
\begin{equation}
\hat P^I_{MK} = {2I+1 \over 16\pi^2} \int d\Omega D^I_{MK}(\Omega)
\hat R(\Omega),
\end{equation}
$\hat R(\Omega)= e^{-\imath \alpha \hat J_z}e^{-\imath \beta
\hat J_y}e^{-\imath \gamma \hat J_z}$ being the rotation operator and
$D^I_{MK}(\Omega)=<\nu IM|\hat R(\Omega)|\nu IK>^*$ its irreducible
representation where $\{|\nu IM>\}$ is a complete set of states for the
specified angular momentum quantum number $IM$. Since the spectral
representation of the projection operator Eq. (7) is represented by
\begin{equation}
\hat P^I_{MK} = \sum_\nu |\nu IM><\nu IK|,
\end{equation}
it is easy to see that $|\Phi'> \equiv e^{-\imath \frac{\pi}{2}\hat
J_z}|\Phi>$, i.e. the state of the opposite triaxiality to a state
$|\Phi>$, is projected to give
\begin{equation}
\hat P^I_{MK}|\Phi'> = \hat P^I_{MK}e^{-\imath \frac{\pi}{2}\hat
J_z}|\Phi> = (-)^{-\imath \frac{\pi}{2}K}\hat P^I_{MK}|\Phi>.
\end{equation}
This state differs only by a phase factor from $\hat P^I_{MK}|\Phi>$ and
thus represents the same physical state. It therefore proves that the
result of the angular momentum projection should be independent of the
sign of $\epsilon'$. We have used this property to check the programming
since it is a non-trivial relation. Note that this justifies the
above-mentioned restriction $\epsilon' \ge 0$.
Details of the projection technique and algorithm are discussed in an
Appendix of \cite{hs95}.

In the present work, we have diagonalized the Hamiltonian Eq. (1) within
the space spanned by $\{\hat P^I_{MK}|\Phi>\}$ where $|\Phi>$ is the
(triaxial) quasiparticle vacuum state. The TPSM eigenvalue equation with
the eigenvalue $E^I$ for a given spin $I$ thus becomes
\begin{equation}
\sum_{K'} \left(H^I_{KK'} - E^I N^I_{KK'}\right)F^I_{K'}=0
\end{equation}
where the matrix elements are defined by
\begin{equation}
H^I_{KK'}=<\Phi|\hat H\hat P^I_{KK'}|\Phi>,~~~
N^I_{KK'}=<\Phi|\hat P^I_{KK'}|\Phi>.
\end{equation}
This TPSM equation has been solved for a range of nuclei in the
rare-earth region and the results of a selected few are presented in
Figs. 1 and 2.

The deformation parameters $\epsilon$ used in Fig. 1 are
exactly the same as those used in the earlier calculations with the
axially symmetric basis \cite{hs95}, i.e. $\epsilon$ = 0.20, 0.20 and
0.225 for $^{156}$Er, $^{158}$Yb and $^{176}$W, respectively. The
results with $\epsilon'=0.0$ in Figs. 1 and 2 represent these axially
symmetric calculations. The experimental moments of inertia (represented
by circles) increase very steeply. The calculations with $\epsilon'=0.0$
on the other hand depict a very slow increase and is typical of an
axially deformed rotational band. The moments of inertia in Fig. 1
become steeper with increasing value of $\epsilon'$ and the value
close to $\epsilon'=0.15$ reproduces the experimental data. Roughly
speaking, this $\epsilon'$ value corresponds to $\gamma=35^0$. It should
be noted that the experimental moment of inertia shown in Fig. 1
slightly increases, in particular for $^{176}$W, at the higher end,
whereas the theoretical moment of inertia shows a drop. This increase in
the observed moment of inertia can be explained by noting that, at
around spin I=$12^+$, a 2-quasiparticle band (i.e. the s-band) will
cross with the ground band and the energy of the higher spin states will
be depressed, so that the moment of inertia will effectively increase.
In the present calculations, the projection has been carried out only
from the ground (i.e. the 0-quasiparticle) band and this effect is not
taken into account. The projection from 2- and higher-quasiparticle
states requires further work and will be reported elsewhere.

Fig. 2 shows the moments of inertia for some Os-isotopes. It is known
that these isotopes are $\gamma$ soft with very low-lying $\gamma$
bands. It is clear from Fig. 2 that, for $^{184}$Os, the moment of
inertia is well reproduced with $\epsilon'=0.15$. For $^{186}$Os and
$^{188}$Os, the experimental moment of inertia can be explained with
$\epsilon'$ between 0.10 and 0.15.

In summary, it has been clearly shown in the present work that
three-dimensional angular momentum projection from triaxial Nilsson+BCS
deformed intrinsic wavefunction is essential for an accurate description
of the transitional nuclei. The moments of inertia of these transitional
nuclei depict a steep increase as a function of rotational frequency
in the low spin region and this can only be explained with triaxial
deformation of $\gamma \simeq 30^0$ since the inclusion of 2- and
higher-quasiparticle bands will affect little in this spin region.

We would like to mention that the present work has been exploratory. For
a detailed study, the energy-surface needs to be analyzed as a function
of $\epsilon$ and $\epsilon'$ to look for the optimal deformation. In
the present work, the deformation parameter $\epsilon$ has been taken
from the earlier studies in which the axial symmetry was assumed. In a
more consistent treatment, both $\epsilon$ and $\epsilon'$ have to be
varied in order to search the energy minimum for spin $I=0$ \cite{hhr84}.

\begin{figure}
\caption{
The experimental and the calculated moments of inertia $(\Theta)$ 
are plotted as a function of the rotational frequency $(\omega)$ for
$^{156}$Er, $^{158}$Yb and $^{176}$W. It is very evident that the Expt.
moment of inertia is reproduced with the triaxiality $\epsilon' \simeq
0.15$ for all the three cases.  
}
\label{figure.1}
\end{figure}

\begin{figure}
\caption{
The experimental and the calculated moments of inertia $(\Theta)$ 
are plotted as a function of the rotational frequency $(\omega)$ for
$^{184}$Os, $^{186}$Os and $^{188}$Os. For $^{184}$Os, the Expt. moment
of inertia is very well reproduced with $\epsilon' \simeq 0.15$. For 
$^{186}$Os and $^{188}$Os, it can be described with $\epsilon'$ between
0.10 and 0.15. 
}
\label{figure.2}
\end{figure}

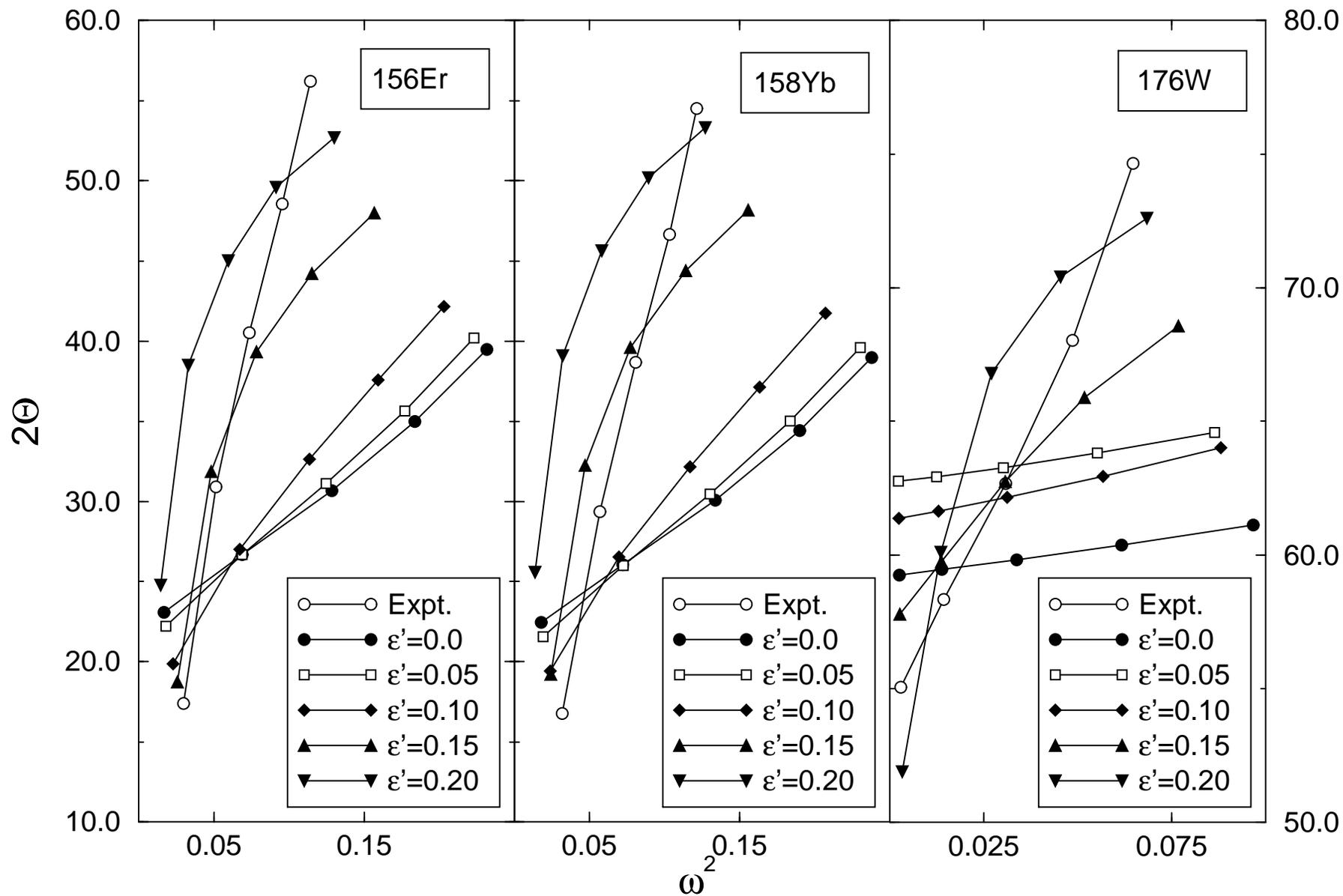
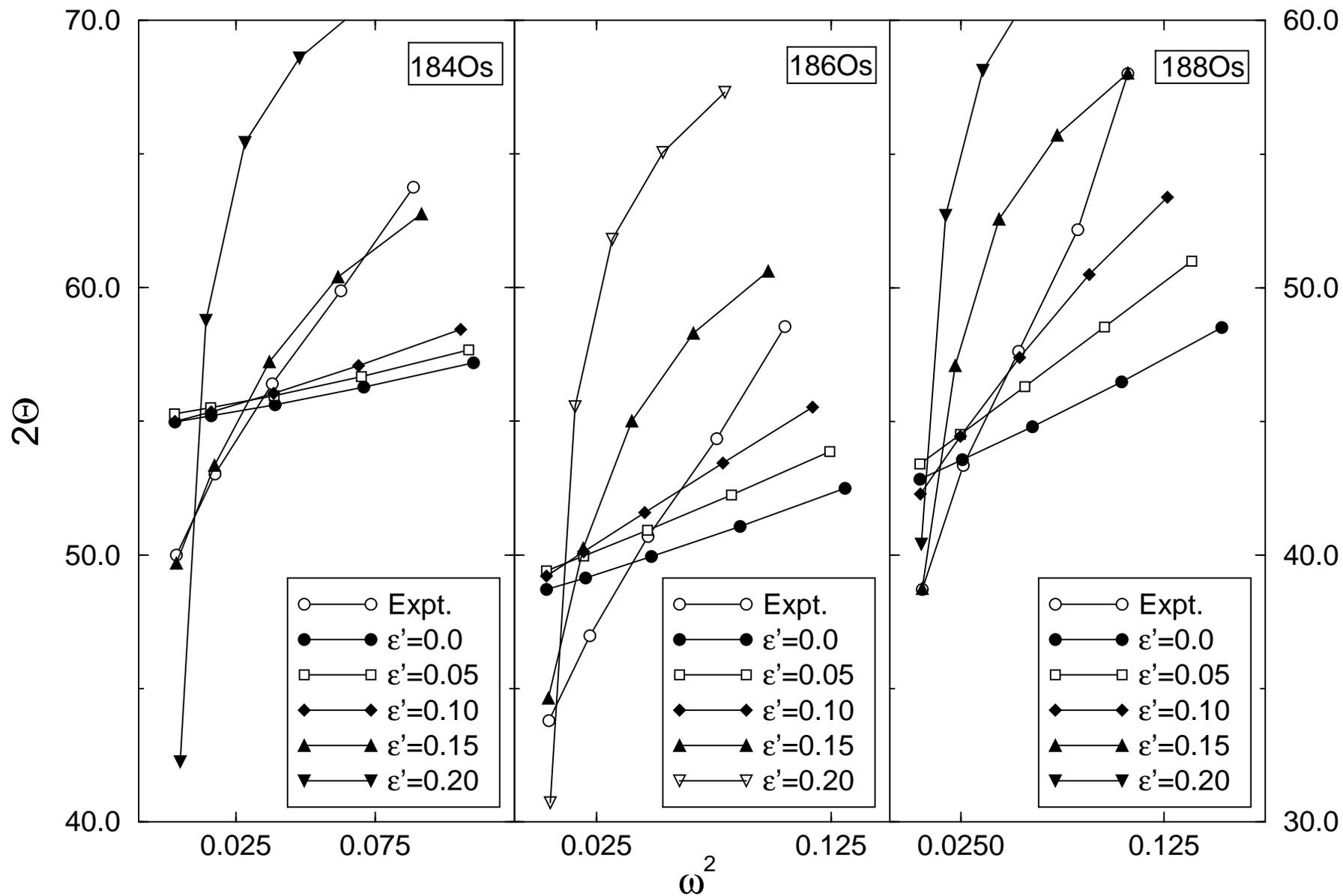
\end{document}